\documentclass[prl,twocolumn,superscriptaddress,showpacs]{revtex4}
\usepackage{color}
\usepackage{amssymb}
\usepackage{upgreek}
\usepackage{graphicx}
\usepackage{mathrsfs}
\usepackage[utf8]{inputenc}
\usepackage{amsmath}

\newcommand{\Nac}{N_{\text{ac}}}
\newcommand{\fsh}{f_{\text{sh}}}
\newcommand{\Hsh}{H_{\text{sh}}}

\newcommand{\kHz}{~\text{kHz}}
\newcommand{\Oe}{~\text{Oe}}
\newcommand{\K}{~\text{K}}

\begin{document}

\title{Ac dynamic reorganization and non-equilibrium phase transition in driven vortex matter}

\author{M. Marziali Berm\'udez}
\affiliation{Departamento de Física, Facultad de Ciencias Exactas y Naturales, Universidad de Buenos Aires, Argentina.}
\affiliation{Instituto de Física de Líquidos y Sistemas Biológicos (IFLYSIB), UNLP-CONICET, La Plata, Argentina}

\author{L. F. Cugliandolo}
\affiliation{Sorbonne Universit\'e, Laboratoire de Physique Th\'eorique et Hautes Energies, CNRS UMR 7589,
4, Place Jussieu, Tour 13,
5\`eme \'etage, 75252 Paris Cedex 05, France}

\author{G. Pasquini}
\email{pasquini@df.uba.ar}
\affiliation{Departamento de Física, Facultad de Ciencias Exactas y Naturales, Universidad de Buenos Aires, Argentina.}
\affiliation{Instituto de Física de Buenos Aires (IFIBA), UBA-CONICET, Buenos Aires, Argentina.}

\date{\today }


\begin{abstract}
Externally driven glassy systems may undergo non-equilibrium phase transitions (NEPTs). In particular, ac-driven systems display special features, like those observed in the vortex matter of $\mathrm{NbSe_2}$, where oscillatory drives reorganize the system into partially ordered vortex lattices. We provide experimental evidence for this dynamic reorganization and we show an unambiguous signature of its connection with an NEPT driven by ac forces. We perform a scaling analysis and we estimate critical exponents for this transition. Our results share similarities with some glass-to-viscous-liquid NEPTs and invite to search for similar physics in other elastic disordered media. 
\end{abstract}

\maketitle


Glassiness is a synonym of rich dynamics, closely related to metastability and history effects in which plasticity may play a key role. 
Indeed, when externally driven, glassy systems may adopt self-organized configurations and undergo non-equilibrium phase transitions~\cite{Hinrichsen2000, Cugliandolo2013, Reichhardt2017}. 
Vortex matter in type-II superconductors shows aspects of glassiness. 
These systems are often modelled as disordered elastic media, a category which also includes magnetic~\cite{Metaxas2007, Jeudy2016} and  ferroelectric~\cite{Tybell2002, Ziegler2013} domain walls, as well as interacting particles in disordered substrates such as colloidal systems~\cite{Reichhardt2017,  Corte2008, Reichhardt2009}, Wigner crystals~\cite{Reichhardt2017, Cha1994}  or skyrmion lattices~\cite{Reichhardt2017, Nagaosa2013}. 
Although the microscopic equations behind these systems are completely different, under some reliable assumptions all of them can be described as elastic manifolds in a disordered landscape \cite{Blatter1994, Kolton2006, Giammarchi2009}. 
In all these cases, a depinning occurs when an external continuous drive is increased beyond a critical value. 
The nature of this transition and its relation to the proliferation or annealing of topological defects have been thoroughly studied, though many questions remain still unsolved in this field~\cite{Reichhardt2017}.  
The dynamics of glassy systems driven by alternating (ac) forces have received much less attention. 
Work carried out mostly during the last decades showed that ac driven systems display special features, not directly translatable from the corresponding dc regimes, and may be dynamically reorganized into different configurations~\cite{Chakrabarti1999, valenzuela2000, valenzuela2001, pasquini2008, daroca2010, daroca2011, marziali2015, domenichini2019}. 
In particular,  molecular-dynamics simulations of ac-driven 2D vortex lattices (VL) revealed a plastic ac depinning in the low-frequency regime for driving ac Lorentz forces $F_\text{L} > F_\text{c}$, which move vortices over distances larger than the typical pinning radius in each ac cycle~\cite{daroca2010}.  
For $F_\text{L}$ between the low-amplitude linear regime and the depinning amplitude $F_\text{c}$, plastic random displacement produces a huge number of VL dislocations, but most vortices remain trapped around the pinning sites and the final configuration depends strongly on the initial conditions. 
On the contrary, above $F_\text{c}$, the memory of the initial configuration is lost after a transient number of cycles ($N_\text{ac}$) that depends on the amplitude and frequency of $F_\text{L}$. 
Past the transient, the density of VL dislocations and the mean vortex velocity remain fluctuating around stationary values which may still depend on the parameters of the ac drive~\cite{daroca2010}. 
These dynamic steady states are reminiscent of the ``fluctuating steady states" observed in colloidal systems~\cite{Corte2008, Reichhardt2009}. 

Driven 2D VL can be used as simple models for a small portion of the vortex systems studied experimentally, for which
history effects in the response have been often observed \cite{daroca2011} and signatures of criticality at the depinning transition were reported~\cite{Reichhardt2017, Kawamura2017}.
In fact, in several superconducting materials, the vortex response is modified after the application of large shaking ac fields and/or transport currents.  Prototypes of such systems are clean $\mathrm{NbSe_2}$ single crystals in which the stable vortex phase at low temperatures and weak magnetic fields is an ordered Bragg Glass (BG) without VL dislocations~\cite{Giammarchi1995}. When field-cooled (FC) from the normal state, the system is trapped in disordered metastable configurations where the VL is strongly pinned~\cite{Yaron1995, marziali2015}. However, by applying high transport current densities~\cite{Yaron1995, Henderson1996, Paltiel2000} or large oscillatory shaking magnetic fields \cite{pasquini2008, daroca2011, marziali2015}, the system overcomes energy barriers and reaches the ordered BG with lower effective pinning. With increasing field and/or temperature, the system undergoes an order-disorder transition to a disordered glass, whose fingerprint is the anomaly known as Peak Effect (PE)~\cite{DeSORBO1964, Larkin1979, Troyanovski2002}, consisting in a sudden increase of the effective pinning.  Intermediate responses that broad the PE have been ascribed in part to surface contamination induced by the probing transport current \cite{Paltiel2000}. However, combined ac susceptibility and small-angle neutron scattering experiments support the existence of a narrow transitional region between the ordered and the disordered phases~\cite{Menon2012, Xiao2000, pasquini2008}, where the application of shaking magnetic fields gives rise to bulk VL configurations with intermediate degrees of disorder, correlated with intermediate~vortex responses~\cite{marziali2015}. A consistent scenario also emerges from tunneling spectroscopy (STS) experiments carried out in the transitional region of Co-doped  $\mathrm{NbSe_2}$ single crystals~\cite{Ganguli2015}. 

In this work, we present experimental results that shows that these “intermediate” configurations are originated from a VL reorganization driven by the oscillatory dynamics. Moreover, unambiguous signatures of criticality suggest that this reorganization is closely associated with a  dynamic phase transition, possibly related with the ac vortex depinning. 

The vortex response may be accessed in different ways. The best choice, given the purpose of the present work (like in Ref. \citep{daroca2011, marziali2015}), is to record the linear ac susceptibility $\chi'$ with a non-invasive measurement. 
In our experiments, this is achieved with the setup and procedure sketched in  Fig.~1a:  a permanent dc field $\mathbf{H}_\text{dc}$ applied on a superconducting single crystal generates a vortex arrangement, which is prepared in an initial  (history-dependent) configuration. At selected times, the ac driving field, that we call shaking field, $\mathbf{H}_{\text{sh}}$,  is switched on,  as a way to reorganize the vortex configuration. Subsequently, the shaking field is switched off. Before and after shaking the system,  the linear response is measured by applying a  very small ac field $\mathbf{H}_{\text{ac}}$. that forces vortices to perform small (harmonic) oscillations inside their effective pinning potential wells, without modifying their spatial configuration. These oscillations propagate through the sample due to the vortex-vortex repulsion, with a characteristic penetration depth $\lambda _{\text{ac}}$ that,  for a fixed experimental geometry, determines the linear ac susceptibility \cite{vanderbeek1993}. In the low frequency Campbell regime \cite{Campbell1971}, vortices oscillate, in a mean-field approximation~\cite{Raes2014}, in phase with $\mathbf{H}_{\text{ac}}(t)$. In this case, $\lambda _{\text{ac}}$ is related with the curvature of the effective pinning potential and determines the in-phase inductive component of $\chi'$, that is nearly frequency independent.  

We used a clean  $1\times 1\times 0.2~\text{mm}^3$ $\mathrm{NbSe_2}$ single crystal with $T_\text{c} = 7.2$~K grown in Bell Labs~\cite{Oglesby1994}. The phase diagram for several crystals from this source
was characterized using a 7-T MPMS XL (Quantum Design)~\cite{marziali2015}. Our crystals only showed minor, insignificant variations between them and compared to other samples reported in the literature. Linear ac susceptibility measurements were done using a homemade susceptometer based on the mutual inductance technique, installed in a cryostat that allows temperature regulation within $\Delta T\leq 2$~mK. In our setup, the dc field ($H_\text{dc} = 1$~kOe) is parallel to both shaking and perturbation fields and the $c$ axis of the sample (Fig.~1a). The latter field has amplitude $H_\text{ac} = 10$~mOe.  With these parameters, linear response, low dissipation, and frequency independence, which are characteristic features of the linear Campbell regime, were verified. We then chose to use $f_\text{ac}=90$~kHz to have a good resolution.
  
Shaking fields,  $H_\text{sh} (t)$, are also applied parallelly to the applied dc field, by controlling the number, amplitude and period of current pulse trains applied to the primary circuit. Local self-heating induced by the dissipation of ac shaking current inside the sample at high shaking amplitudes and/or frequencies was avoided by applying temporally spaced short bursts. Because the order-disorder transition region is characterized by a sharp increase in the ratio between effective pinning and elastic forces with temperature,  the system's behavior is strongly temperature-dependent. Therefore, good control over the temperature in the transitional PE region was assured.

\begin{figure}[t]
\centering
\includegraphics[width=85mm]{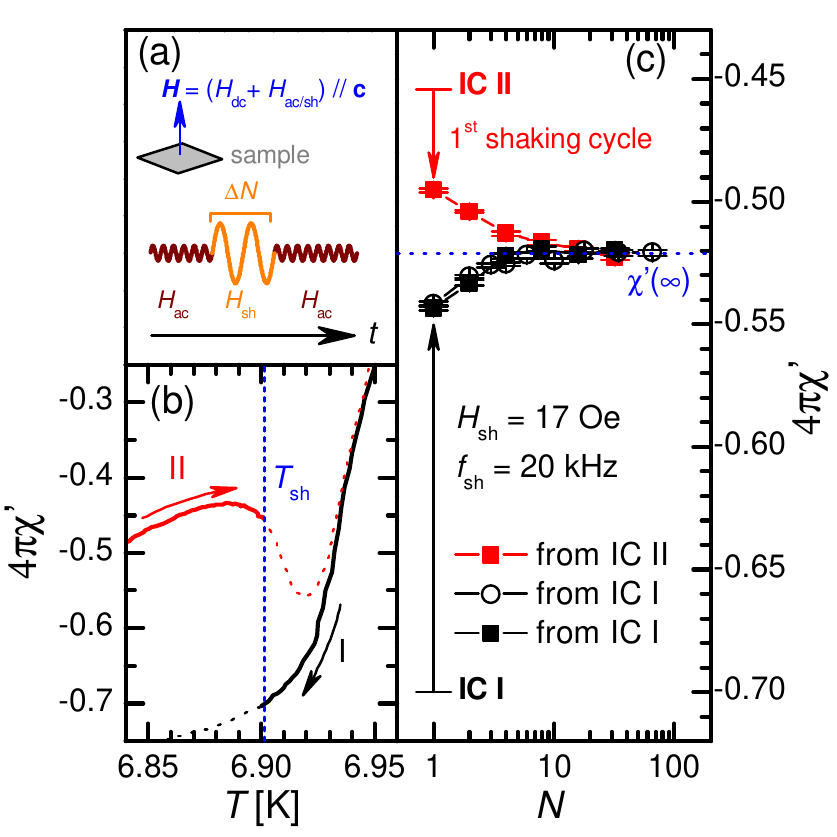}
\label{fig1}
\caption{ (Color online.) (a) Sketch of the experimental setup: the  dc field ($\mathbf{H}_\text{dc}$) is parallel to both shaking ($\mathbf{H}_\text{sh}$) and perturbation ($\mathbf{H}_\text{ac}$) fields, and points along the c axis of the sample. At selected times, the $\chi'$ measurement is interrupted to apply a burst of $\Delta N$ shaking pulses.  (b)  History dependent in-phase component of $\chi(T)$ in the transitional PE region, corresponding to FC disordered (black) and warming ordered (red) VL configurations (see text). 
(c) Evolution of $\chi'$, after applying $N$ shaking pulses starting from the IC I (black dots and squares) and IC II (red squares)
stabilized at temperature $T_\text{sh}$ (vertical dotted line in (b)). After a transient number of cycles $N_\text{ac}$, an IC-independent value $\chi'_\infty$ is reached.
}
\end{figure}

Figure 1b displays different linear ac susceptibility responses in the transitional region. 
The black curve shows the response measured during a FC procedure, where the VL is disordered and strongly pinned. 
The red curve displays the response of an ordered, weakly pinned VL, recorded in a warming procedure, after having shaken the system with $1000$ cycles of a large ac field ($24\Oe$, $1\kHz$) at low temperature ($\sim \! 6.5\K$). 
These responses, well defined and repeatable under the same cooling/warming protocols, can be used as different \textit{Initial conditions} (IC) for our experiments.  
Condition I corresponds to the disordered FC VL and condition II to the ordered VL. In absence of any additional driving force, there is no significant evolution of the susceptibility with time. 
A small relaxation is observed when starting from condition II (not discussed here) but the system remains mainly trapped near these initial configurations. 

However, as observed in other complex systems, like granular matter~\cite{Jaeger1996, Berthier2001}, for example, large oscillatory driving forces may dynamically assist the system to evolve towards a stationary state, independent of the initial condition. 
We move then to the proper dynamic measurements of our interest.
Once an IC is selected, a temperature in the transitional region is chosen and made stable to ensure that competing interactions are not modified during the experiment. 
We then apply a series of pulses of amplitude $H_\text{sh}$ and period $1/f_\text{sh}$, and we measure $\chi$ after $1, \, 2, \dots, \, N$ pulses.  
Its resulting in-phase component as a function of the number of pulses, $N$, is shown in  Fig.~1c. 
Red squares show its evolution starting from an initial condition in the branch II (see Fig. 1b), whereas black dots and squares show two examples of the evolution starting from a condition I. 
The effect of the first pulse dependends strongly on the initial condition. 
However, after a transient number of cycles that we call $N_\text{ac}$, the responses converge to a common stationary value. 
We have checked that the evolution, $\chi'(N)$, does not depend on the number of pulses conforming each applied shaking burst and that it is not modified by the very weak perturbation used during the measurements.

Both the evolution and stationary response may depend on the amplitude, $H_\text{sh}$, and frequency, $f_\text{sh}$, of the shaking field. The variation of $\chi'(N\rightarrow \infty )= \chi'_{\infty }$, as a function of  $H_\text{sh}$,  is displayed, for selected $f_\text{sh}$, in the inset of Fig.~2c.  On the one hand, the asymptotic value increases with $f_\text{sh}$ at frequencies higher than $3$ kHz but is independent of the shaking frequencies at low $f_\text{sh}\lesssim 1$~kHz \cite{daroca2011}. On the other hand, at a fixed $f_\text{sh}$, the final response is nearly  $H_\text{sh}$-independent, as well as independent on the initial condition, for  $H_\text{sh} > 1$~Oe, although it keeps its frequency dependence even for these large field amplitudes. Therefore, quite generally, the driven vortex system evolves to different dynamically organized stationary configurations. 
Whether the response easily converges to a final value independent of the initial conditions, depends on the shaking amplitude $H_\text{sh}$, which determines the typical vortex displacements.

\begin{figure}[tb]
\centering
\includegraphics[width=85mm]{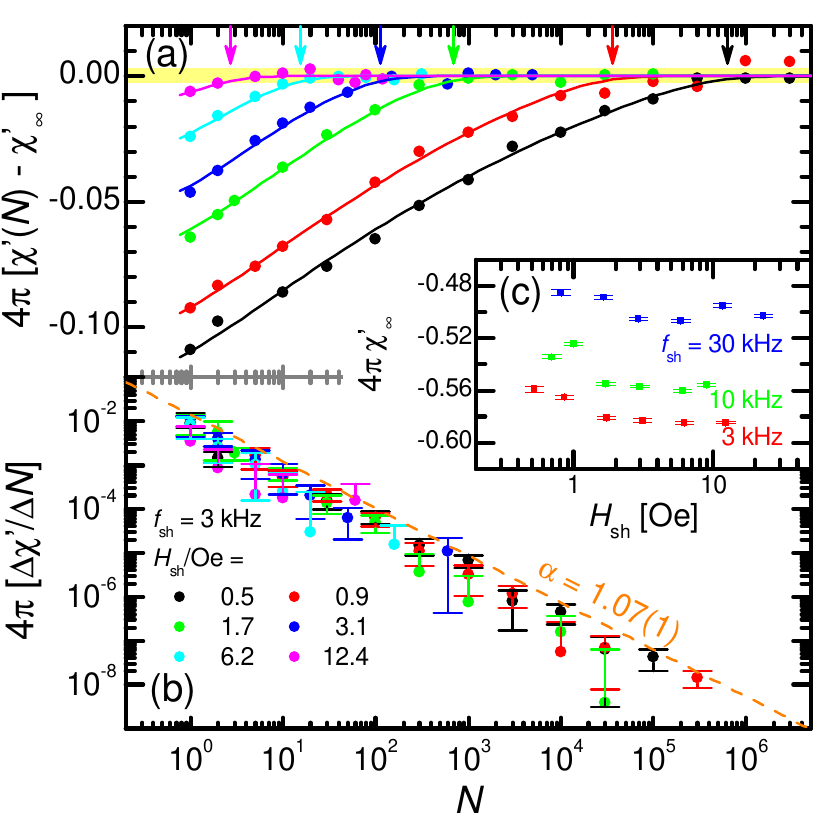}
\label{fig2} 
\caption{(Color online.) (a) Difference between the in-phase linear response after $N$ shaking pulses, $\chi'(N)$, and its asymptotic stationary value, $\chi'_{\infty }$, for different amplitudes of the shaking field, $H_\text{sh}$, at $f_\text{sh} = 3$~kHz (log-linear scale). Continuous lines are the functions in Eq.~(\ref{eq:continuous lines}) and 
the vertical arrows indicate the characteristic number of cycles $\Nac$ (its parameter dependence is shown in the main panel in Fig.~\ref{fig4}).
(b) Finite difference approximation to the derivative of the functions above, against $N$, in double logarithmic scale. 
The slope of the continuous orange line yields the critical exponent $\alpha$. 
(c) Stationary linear response, $\chi'_{\infty }$, as a function of the shaking amplitude $\Hsh$ for several  $\fsh$.
}
\end{figure}

Figure~2a shows an example of the evolution of $\chi'$ with the number of applied pulses $N$ tending to the stationary response $\chi'_{\infty }$, for different amplitudes ranging from $0.5$~ Oe to $12$~Oe, at $\fsh = 3$~kHz. 
A quick and good convergence to $\chi'_\infty$ in a characteristic number of cycles $N_\text{ac}\lesssim 1000$ is achieved for $H_\text{sh}\gtrsim 1$ Oe, that completely penetrates the sample,  for which we can estimate a vortex displacement larger than the VL parameter in most of the sample (see SM in Ref. \cite{marziali2015}). 
However, with decreasing amplitudes, $N_\text{ac}$ increases by several orders of magnitude, beyond the experimentally accessible range, suggesting a divergence. A possible origin for this divergence could be the fact that the shaking field plays the role of an ``effective temperature'' in an activated process, where activation barriers grow when approaching the stable state \cite{Louden2019}.  We have tested that our data are not compatible with the expected behavior \cite{SM} and we therefore discarded this possibility. On the other hand, a divergent $N_\text{ac}$ at a critical shaking amplitude $H^c_\text{sh}$ could be indicative of a dynamic phase transition. We explore the latter possibility in the following.

In case we were confronted to a dynamic phase transition, a critical behavior should be observed near the critical field, 
where the characteristic time $\tau$ diverges; in a dc driven system, we should then expect~\cite{Reichhardt2009, Reichhardt2017}
\begin{equation}
\langle x(t) \rangle 
\propto \frac{e^{-t/\tau }}{t^{\alpha }}\sim \left\{ \begin{array}{lcc}
             t^{-\alpha } &  \;\; \mbox{if} \;\; & t \ll \tau \; ,  \\
             \\e^{-t/\tau} &  \;\; \mbox{if} \;\; & t > \tau \; , 
             \end{array}
   \right.
\end{equation}
for any observable coupled to the order parameter, that tends to $0$ under the stationary condition. 

In the ac driven vortex system, we propose a similar critical behavior with time measured by the number of applied cycles $N$. We chose as our observable the rate of change of the linear in-phase ac susceptibility per shaking cycle.
Hence, we expect an evolution of the form 
\begin{equation}
4\pi \, (\chi'(N) - \chi'(N-1)) =A \, \frac{e^{-N/N_\text{ac}(H_\text{sh})}}{%
N^{\alpha }}
\; , 
\label{eq:criticallity}
\end{equation}
which implies
\begin{eqnarray}
\frac{4\pi }{A}(\chi'_{\infty }-\chi'(N)) =\sum_{k=N+1}^{\infty}\frac{e^{-k/N_\text{ac}}}{k^{\alpha }} \qquad\qquad\qquad &&
\label{eq:continuous lines}
\nonumber\\
\qquad
\simeq\int_{N+1/2}^{\infty } \!\! dk \, \frac{e^{-k/N_\text{ac}}}{k^{\alpha }} = N_\text{ac}^{1-\alpha }\Gamma _{1-\alpha }\left( \frac{N+1/2}{N_\text{ac}}\right) 
& & \label{eq:analitycal}
\end{eqnarray}
that in turn has been approximated, in the continuous limit, 
by the incomplete Gamma function, $\Gamma _{\nu}(z)= \int_{z}^{\infty }t^{\nu -1}e^{-t}dt $. 
Here, shifting the lower limit down by $1/2$ compensates for truncation errors (see SM for more details).

Figure 2b shows the evolution of $\Delta \chi'/\Delta N$ with $N$ obtained from the data shown in Fig.~2a, in double logarithmic scale; the dashed line indicates the expected power-law behavior for the exponent $\alpha =1.07\pm 0.01$, which best fits data for this particular frequency. Continuous lines in Fig.~2a are fits of the data with the function (\ref{eq:continuous lines}) using a unique exponent $\alpha$ and prefactor $A$ but different $N_\text{ac}(H_\text{sh})$ \cite{SM}, indicated by vertical arrows. 
 
\begin{figure}[b]
\centering
\includegraphics[width=85mm]{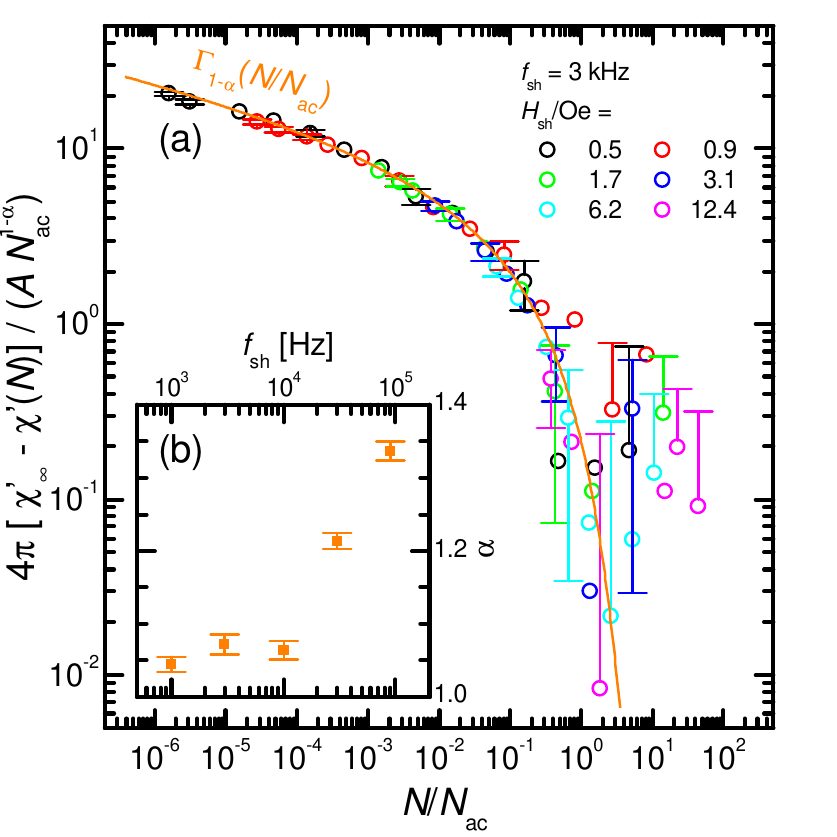}
\label{fig3}
\caption{(Color online.) (a) Scaling of the data shown in Fig.~2. The master curve is well 
represented by $\Gamma_{1-\alpha} (N/N_\text{ac})$, plotted with a continuous orange line. 
(b) The exponent $\alpha$ vs. $f_\text{sh}$.
}
\end{figure}

Equation~(\ref{eq:analitycal}) implies that, after normalisation by $\Nac^{1-\alpha}$, 
the data should  scale when plotted as a function of $N/\Nac$, 
for $N\gg 1$.
Such scaling is shown in Fig. 3, where all data sets used in Fig. 2 are plotted. Data collapse is excellent up to $N/\Nac\simeq 1$, 
and  the master curve $\Gamma_{1-\alpha}(N/N_\text{ac})$ (continuous line) represents the data very accurately. A similarly good scaling is obtained for shaking frequencies ranging from $f_\text{sh} = 1$ kHz to $90$ kHz \cite{SM}. 

The critical exponent  $\alpha$ is expected to be unique for each universality class of phase transitions. The resulting exponents $\alpha(f_\text{sh})$ are plotted in the inset of Fig. 3. A single frequency-independent $\alpha =1.02\pm 0.02$ is obtained for $f_\text{sh}$ up to $10$ kHz, but $\alpha$ increases at higher frequencies. 

The characteristic number of cycles $N_\text{ac}$ (indicated by vertical arrows in the example of Fig.~2b) is expected to diverge at a frequency dependent critical field $H_\text{sh}^c(f_\text{sh}$) as 
\begin{equation}
N_\text{ac}(H_\text{sh},f_\text{sh})=N_{0} \; (H_\text{sh}-H_\text{sh}^c.(f_\text{sh}))^{-\beta }
\; , 
\label{eq:NAC}
\end{equation}
with $\beta$ another critical exponent.  By assigning weights to the expected models (Eqs.~(\ref{eq:continuous lines}) and (\ref{eq:NAC})), we were able to fit the whole set of data for all the shaking amplitudes and frequencies between $1$ and $30$ kHz, obtaining a single critical exponents $\beta$ \cite{SM}. The main panel in Fig.~4 shows the resulting transient number of cycles $N_\text{ac}$ as a function of $H_\text{sh}-H_\text{sh}^c(f_\text{sh})$ for different $f_\text{sh}$. All data collapse on a linear relationship with  $\beta =2.4\pm 0.4$ (dashed orange line), supporting the existence of an ac dynamic transition. The inset in Fig.~4 shows the resulting shaking critical field $H_\text{sh}^c$ as a function of  $f_\text{sh}$. 

\begin{figure}[tb]
\centering
\includegraphics[width=85mm]{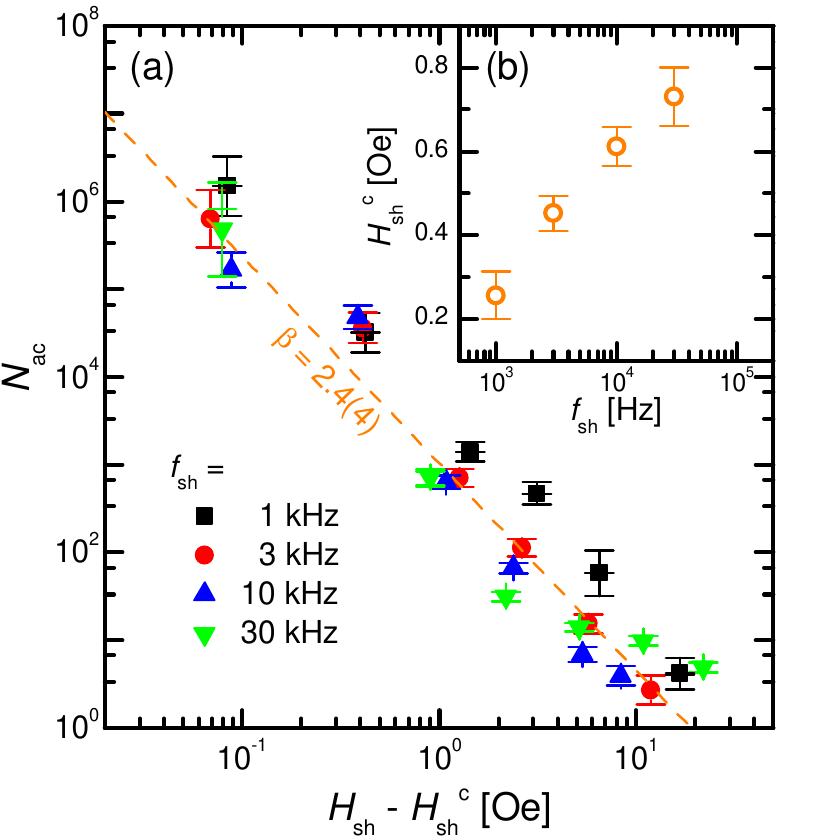}
\caption{(Color online.)  (a) Dependence of the characteristic $N_\text{ac}$ with the distance from the 
critical field amplitude $H_\text{sh}-H_\text{sh}^c(f_\text{sh})$ for four frequencies given in the key, cfr. Eq.~(\ref{eq:NAC}), 
in double logarithmic scale. The exponent $\beta$ extracted from a fit is given in the plot.  
(b) The critical shaking field amplitude, $H_\text{sh}^c$, as a function of shaking frequency.
}
\label{fig4}
\end{figure}

A similar dynamic phase transition has been found in a mean-field disordered model driven by an ac field used to model shaken granular systems~\cite{Berthier2001}. In this simple spin model, the linear response $R$ converges in a finite transient to a stationary form whenever the system is assisted with strong enough ac fields. The transient time (measured in number of cycles) grows with decreasing amplitudes and diverges at a frequency dependent critical amplitude $H_\text{sh}^c$, similarly to what is reported in Fig. 4. This transition is, therefore, between a viscous liquid and a glassy phase. The former is the steady-state phase for field amplitudes above $H_\text{sh}^c(f_\text{sh})$. 

In the present case, reaching the memoryless steady vortex state becomes more and more difficult when the strength of the shaking field decreases until it can no longer do it below the critical curve $H_\text{sh}^c(f_\text{sh})$. A rough estimate of the vortex displacements during each shaking cycle indicates, under shaking fields of the order of the measured $H_\text{sh}^c(f_\text{sh})$, that vortices
move over distances that are smaller than the VL  parameter and (on average) of the order of the coherence length (i.e. the typical pinning radius). 
Therefore, at lower shaking amplitudes, below the ac depinning, the system remains glassy. On the other hand, beyond the ac depinning, vortices move and reorganize into dynamic stationary states which lead to partially ordered vortex configurations.  Whether this dynamic phase is characterized by steady fluctuating states \cite{Corte2008}, as suggested by 2D molecular dynamic simulations \cite{daroca2011}, is still an open question. 

Although checking universal properties is notably difficult in dynamic phase transitions, we have succeeded in using critical slowing down and scaling arguments to characterize $\Delta \chi'/\Delta N$ and the deviation of $\chi'(N)$ from the asymptotic value $\chi'_{\infty}(H_\text{sh}, f_\text{sh}$), respectively.  We found parameter independent values of the critical exponents $\alpha$ and $\beta$, as expected, for low $f_\text{sh}$ and all amplitudes. To the best of our knowledge, this is the first time that a critical behavior, associated with a  non-equilibrium phase transition driven by ac forces, is experimentally reported in vortex matter.

While universal independent values of the critical exponents are expected, deviations from constant $\alpha$ were measured at high $f_\text{sh}$. This observation, together with the increase of the stationary $\chi'_{\infty}(H_\text{sh}, f_\text{sh})$ at high shaking frequencies (characteristic of a more ordered VL configuration), 
suggest a change in the universality class of the phase transition. The influence of viscous losses during shaking could promote a crossover from plastic to elastic dynamics at depinning. This possible explanation, however, deserves further investigation.

In summary, we were able to experimentally observe a dynamic phase transition in a vortex system driven by ac forces, probably associated with ac depinning. The close similarity of vortex matter with other glassy systems opens the perspective to observe similar behavior in systems belonging to the broad class of elastic disordered media.

The authors acknowledge especially useful insights from V. Bekeris and thank  M. Eskildsen, X. S. Ling, M. Mungan, S. Nagel and S. Sastry for interesting discussions. This work was partially supported by the National Scientific and Technical Research Council - Argentina (CONICET) and the University of Buenos Aires. LFC is a member of Institut Universitaire de France and thanks the KITP at UCSB for hospitality.

\bibliography{references.bib}



\end{document}


\title{Supplemetary Material to ``AC dynamic reorganization and non-equilibrium phase transition in driven vortex matter''}
	
\author{M. Marziali Berm\'udez}
\affiliation{Departamento de Física, Facultad de Ciencias Exactas y Naturales, Universidad de Buenos Aires, Argentina.}
\affiliation{Instituto de Física de Líquidos y Sistemas Biológicos (IFLYSIB), UNLP-CONICET, La Plata, Argentina}

\author{L. F. Cugliandolo}
\affiliation{Sorbonne Universit\'e, Laboratoire de Physique Th\'eorique et Hautes Energies, CNRS UMR 7589,
4, Place Jussieu, Tour 13,
5\`eme \'etage, 75252 Paris Cedex 05, France}

\author{G. Pasquini}
\email{pasquini@df.uba.ar}
\affiliation{Departamento de Física, Facultad de Ciencias Exactas y Naturales, Universidad de Buenos Aires, Argentina.}
\affiliation{Instituto de Física de Buenos Aires (IFIBA), UBA-CONICET, Buenos Aires, Argentina.}

	\date{\today }

	\maketitle
	
	\section{A model for the evolution of $\chi'(N)$}
	\subsection{Experimental observables and critical behavior}
	
	As described in the main text, we measured the evolution of the vortex response after the application of $N$ alternating shaking ac pulses, by recording the linear ac susceptibility $\chi'(N)$. We show that, for high enough shaking amplitudes, 
	$H_\text{sh}$, after a transient number of cycles,  $N_\text{ac}$, the response converges to a stationary value, $\chi'_{\infty }$, that is independent of the initial conditions.
	We found that $N_\text{ac}$ increases by several orders of magnitude with decreasing $H_\text{sh}$ and we interpreted this trend as leading to a dynamic critical phase transition. 
	
	An alternative explanation based on activation barriers, where the shaking field plays the role of an ``effective temperature'',  similar to the one described, for example, in Ref. \cite{Louden2019}  has also been tested.  In such a case, the activation barriers can be estimated from the relative rate of change of the susceptibility $(\Delta \chi' /\Delta N)/(\chi' - \chi'_{\infty })$ normalized by the effective temperature. We have tested that our data are not compatible with the expected scaling (Fig.~3 in Ref. \cite{Louden2019}) and we, therefore, discarded this possibility.
	
	If we were confronted with a dynamic phase transition, a critical behavior should be observed near the critical field, $H_\text{sh}^c$.
	Then, for any observable $x(t)$ coupled to the order parameter, which tends to $0$ under the stationary condition, we should expect a time evolution
	of the kind
	\begin{equation}
		\langle x(t) \rangle 
		\propto \frac{e^{-t/\tau }}{t^{\alpha }}\sim \left\{ \begin{array}{lcc}
			t^{-\alpha } &  \;\; \mbox{if} \;\; & t \ll \tau \; ,  \\
			\\e^{-t/\tau} &  \;\; \mbox{if} \;\; & t > \tau \; , 
		\end{array}
		\right.
	\end{equation}
	where $\tau$ is a characteristic time and $\alpha$ is a critical exponent. 
	Furthermore, $\tau$ is expected to diverge as the critical transition is approached, according to some power law with another critical exponent, $\beta$. 
	
	In the system under study, the linear in-phase ac susceptibility $\chi'(N) \rightarrow \chi'_{\infty }$ as $N \rightarrow \infty$ and, consequently, the rate $\Delta \chi' /\Delta N \rightarrow 0$.
	Moreover, as shown in Fig. 2 in the main text, there is a characteristic number of cycles $N_\text{ac}$, such that 
	\begin{equation}
		\frac{\Delta \chi'}{\Delta N } \propto \; N^{-\alpha} \; \text{for} \;\; N \ll N_\text{ac}
		\; ,
	\end{equation}
	with $\alpha$ of order unity.
	It can be readily seen that $\Nac$ grows beyond the experimentally accessible range as the shaking field amplitude, $H_\text{sh}$, is reduced, suggesting a potential divergence
	\begin{equation}
		N_\text{ac} \propto (H_\text{sh} - H_\text{sh}^c)^{-\beta}
		\; ,
	\end{equation}

	The analogy invites us to associate the rate of change of 
	$\chi'$ per shaking cycle as the objective observable, i.e., we propose a model of the form
	\begin{equation}
		\chi'(N) - \chi'(N-1) = \frac{A}{4\pi} \frac{e^{-N/N_\text{ac}}}{N^\alpha}
		\; ,
	\end{equation}
	where $A$ is a constant amplitude factor.
	As a result,
	\begin{equation}
		\begin{split}
			\chi'(N)  = \; & \; \chi'(0) + \frac{A}{4\pi}  \sum_{n=1}^{N} \frac{e^{-n/N_\text{ac}}}{n^\alpha} \\
			= \; & \; \chi'_{\infty } - \frac{A}{4\pi}  \sum_{n=N+1}^{\infty} \frac{e^{-n/N_\text{ac}}}{n^\alpha}.
		\end{split}
		\label{eq:chiSumN}
	\end{equation}
	A critical phase transition would imply
	\begin{equation}
		N_\text{ac}(H_\text{sh},f_\text{sh})=N_{0} \; (H_\text{sh}-H_\text{sh}^c(f_\text{sh}))^{-\beta }
		\; .
		\label{eq:NAC}
	\end{equation}
	
	\subsection{Continuous limit approximation}
	The sum (\ref{eq:chiSumN}) is convergent and, for instance, the value of $ \chi'_{\infty }$ can be calculated exactly for $\alpha = 1$, 
	\begin{equation}
		\chi'_{\infty } = \chi'(0) - \frac{A}{4\pi} ~\log \left(1 - e^{-1/N_\text{ac}} \right).
	\end{equation}
	For other $\alpha$ and finite $N$, one must either  sum all terms explicitly or apply some approximation.
	It is reasonable to assume that, for $N$ large enough, the sum in (\ref{eq:chiSumN}) could be approximated by an integral
	\begin{equation}
		\sum_{n=N+1}^{\infty} \frac{e^{-n/N_\text{ac}}}{n^\alpha} \simeq 
		\int_{N+1/2}^{\infty} dn \; \frac{e^{-n/N_\text{ac}}}{n^\alpha} 
		\; ,
	\end{equation}
	where the lower limit of integration has be shifted down by 1/2 so as to compensate for rounding errors (when compared to the discrete sum).
	By substituting $k = n/N_\text{ac}$, the last integral can be rewritten to match the definition of the incomplete gamma function,
	$\Gamma_\nu(z) = \int_{z}^{\infty} k^{\nu-1} e^{-k} dk$,
	so that
	\begin{equation}
		\chi'(N) \simeq  \chi'_{\infty } - \frac{A}{4\pi}~ N_\text{ac}^{1-\alpha} ~ \Gamma_{1-\alpha}\left(\frac{N+1/2}{N_\text{ac}} \right).
		\label{eq:chiGamma}
	\end{equation}
	\begin{figure}[b]
		\centering
		\includegraphics[width=85mm]{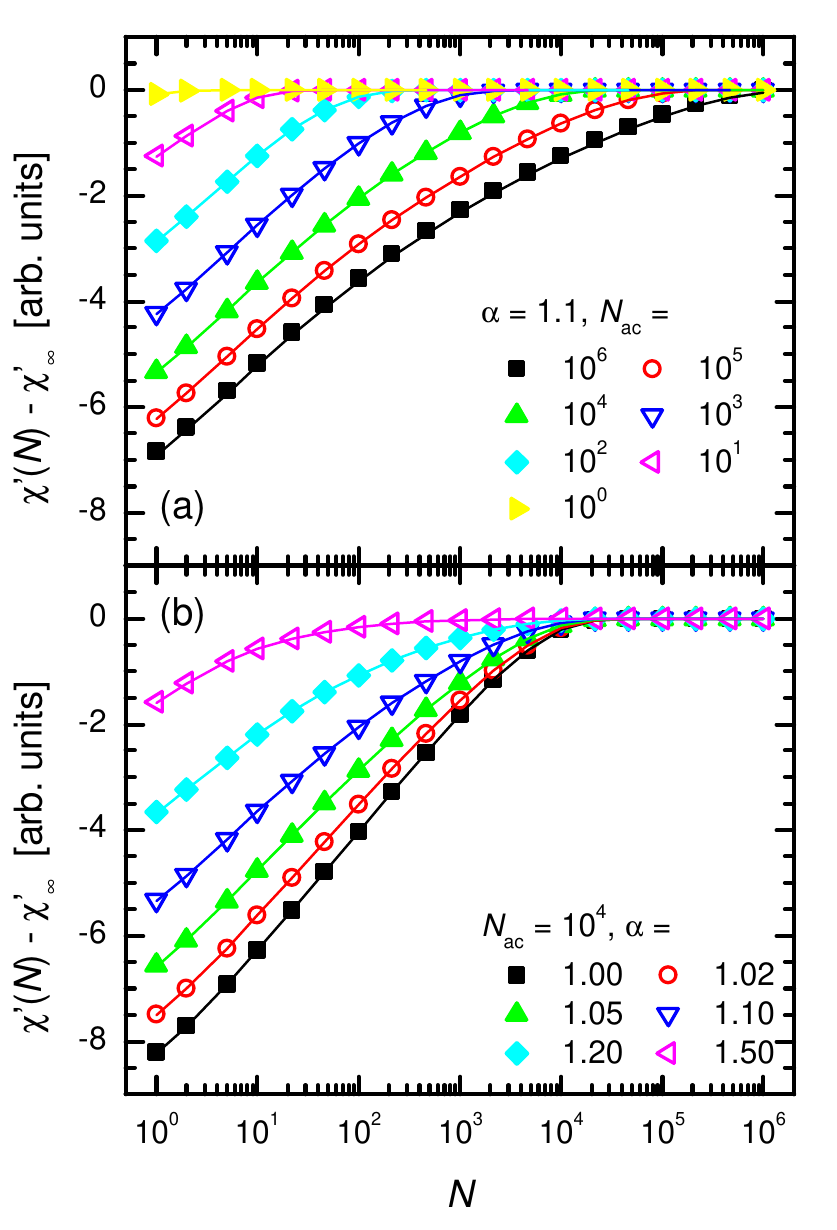}\\
		\caption[S]{Comparison of $\chi'(N)$ between discrete sums (dots, Eq.~(\ref{eq:chiSumN})) and the integral approximation (lines, Eq.~(\ref{eq:chiGamma})) for fixed $\alpha = 1.1$ and various $N_\text{ac}$ (panel a), and fixed $N_\text{ac} = 10^4$ and various $\alpha$ (panel b).}
		\label{fig:GammaSumCompare}
	\end{figure}
	The discrete sums and the integral approximation are compared in Fig.~\ref{fig:GammaSumCompare}. 
	The accuracy of the approximation is remarkable, even for small $N$.

	\section{Procedure used to study criticality}
	
	The procedure described in the following aimed to determine the  exponents, 
	$\alpha$ and $\beta$, and the values of the critical field, $H_\text{sh}^c$, 
	which best fit the experimental data with the expressions (\ref{eq:NAC}) and (\ref{eq:chiGamma}).
	
	\subsection{Preliminary considerations and notation}
	
	The evolution of $\chi'(N)$ has been measured for shaking frequencies $f_\text{sh} = 1$, $3$, $10$, $30$ and 
	$90$ kHz and a broad range of shaking amplitudes.  
	
	The rate $\Delta \chi' /\Delta N$ plotted against $N$ in log-log scale (see, for example, Fig.~2b) 
	yields valuable information about the value of the the $\alpha$ exponent. After a first preliminary analysis  of the whole set of data  across various values 
	of $f_\text{sh}$ and $H_\text{sh}$, we find that a single $\alpha$ seems to describe the data for the whole range of $H_\text{sh}$ 
	at each particular $f_\text{sh}$. However $\alpha$($f_\text{sh} = 90$ kHz) is obviously larger than  $\alpha$($f_\text{sh} = 1$ kHz). 
	Therefore, in our experiment, $\alpha$ cannot be universally determined for all $f_\text{sh}$.
	
	On the other hand, due to the self-heating of the sample, we were not able to measure the evolution of $\chi'(N)$ at high shaking frequencies and amplitudes. Therefore, we have not 
	enough shaking amplitudes  at $f_\text{sh} = 90$ kHz to obtain a reliable 
	$\beta$ exponent for this frequency. Our aim was then to fit all the data from $1$ to $30$ kHz with a single $\beta$ exponent.

	Let the set of indices $j$, $k$, and $l$ denote the $j$-th shaking frequency $f_j \in \{1\khz,3\khz,10\khz,30\khz,$ $90\khz\}$, the $k$-th shaking amplitude $H_{j,k}$ (within the set of amplitudes applied at $f_\text{sh} = f_j$), and the $l$-th accumulated number of cycles $N_{j,k,l}$ in each data series (corresponding to $f_\text{sh} = f_j$ and $H_\text{sh} = H_{j,k}$), respectively.
	Due to the hierarchical structure of data, we will omit indices whenever they are evident from the context to simplify the notation.
	In addition, to avoid confusion with the $\chi^2$ statistic and further clean the notation, we will use $z_l = 4\pi \chi'(N_l)$.
	
	Taking into account the previous considerations,  we will need to determine five different exponents $\alpha_j = \alpha(f_j)$, without any constraint on their functional relationship.
	Consequently, we will let the prefactor $A = A_j$ vary with $f_j$.
	Although it might be possible that $A$ depended on $H_{j, k}$ too,  such freedom in the model was not necessary and would have led to overfitting.
	On the other hand, as shown in Fig.~2c, the asymptotic values $\tilde{z}_{j,k} = 4\pi \chi'_{\infty }$ depend on both $f_j$ and $H_{j,k}$. 
	
	As discussed above, the characteristic number of cycles $N_\text{ac}$ depends on both $f_j$ and $H_{j,k}$. 
	In contrast with previous parameters, a universal power-law behavior with $H_{j,k}$ is expected near the critical shaking amplitude $H_j^c$,  for all $f_j$. Therefore, we include this expected dependence as an additional constraint. Due to the statistical character of the evolution, we assume a statistical dispersion of the experimental $N_\text{ac}$ from the expected critical behavior.  
	
	Finally, for numerical reasons, it is convenient to define $\lambda = \log N_\text{ac}$ and to write the power law in the linear form
	\begin{equation}
		\lambda_{j,k} = \tilde{\lambda}_j - \beta ~ \log(H_{j,k} - H_j^c)
		\; ,
	\end{equation}
	where $\tilde{\lambda}_j$ is an offset term which we will allow to vary with $f_j$.

	\subsection{Joint regression for $\chi'(N)$ and $N_\text{ac}(H_\text{sh})$}
	
	We shall assume normal error distributions for $z_l$ and $\lambda_{j,k}$ (i.e., log-normal errors in $\Nac$) so that the maximum likelihood estimates for the model parameters can be obtained by least square error methods.
	In so doing, the (log-transformed) characteristic number of cycles ($\lambda_{j,k}$) behaves as a parameter in the model for $z_l \sim z(N_l)$ but, at the same time, they are dependent variables in terms of the shaking amplitude. 
	Moreover, since the total number of parameters is quite large (73), straight application of non-linear optimization algorithms may be troublesome.
	Hence, we need careful statistical treatment.
	
	\subsubsection{Divide and conquer}
	First, let     $g_{j,k,l} = g(N_l, \lambda_{j,k}, \alpha_j)$, with
	\begin{equation}
		g(N, \lambda, \alpha) = e^{(1-\alpha)\lambda} ~ \Gamma_{1-\alpha}\left(\frac{N+1/2}{e^\lambda} \right)
		\; .
	\end{equation}
	Now, given $\lambda_{j,k}$ and $\alpha_j$, we can write a linear model for $z_l$,
	\begin{equation}
		z_{j,k,l} = A_j~ g_{j,k,l} + \tilde{z}_{j,k} + \epsilon_{j,k,l}
		\; ,
	\end{equation}
	where 
	$\epsilon_{j,k,l}$ are a set of normally distibuted errors with zero mean and unifrom variance equal to $\sigma_z^2$.
	Similarly, given $H_j^c$, let $h_{j,k} = \log (H_{j,k} - H_j^c)$. Then, we arrive to another lineal model
	\begin{equation}
		\lambda_{j,k} = \tilde{\lambda}_j - \beta ~ h_{j,k} + \xi_{j,k}
		\; ,
	\end{equation}
	where $\xi_{j,k}$ are a set of normally distibuted errors with zero mean and variance equal to $\sigma_\lambda^2 + e^{-2 \lambda_j}$.
	Here, $\sigma_\lambda^2$ is a constant variance which models inherent randomness, while the term $e^{-2 \lambda_j}$ (equivalent to an error of a single cycle in $\Nac$) acts as a weighting factor that allows for larger deviations from the ideal power law as $\Nac$ approaches zero. 
	
	As a result, maximum likelihood estimates for parameters in the set $\mathbb{P}_\text{lin} = \{ \beta, \tilde{\lambda}_{j}, A_j, \tilde{z}_{j,k} \}$, 
	given the values of the remaining parameters, $\mathbb{P}_\text{nl} = \{H_{j}^c, \alpha_j, \lambda_{j,k} \}$,  
	can be derived from the minimization of the conditional log-likelihood function
	\begin{eqnarray}
		&& -2 \log \mathcal{L}_c(\mathbb{P}_\text{lin} | \{z_l\}, \mathbb{P}_\text{nl})  
		\nonumber\\
		&& 
		\qquad\quad
		=
		\sum_{j,k,l}\frac{\epsilon_{j,k,l}^2}{\sigma_z^2} 
		+
		\sum_{j,k}\frac{\xi^2_{j,k}}{\sigma_\lambda^2 + e^{-2 \lambda_j}} 
		\label{eq:logLc}
	\end{eqnarray}
	with the usual weighted linear least squares method.
	Furthermore, minimization can be carried out separately on each $j$-term of the first and second sums.
	Then, the minima of the conditional likelihood function for each combination of the parameters in $\mathbb{P}_\text{nl}$ constitute a profile log-likelihood function for these parameters,
	\begin{eqnarray}
		&& 
		-2 \log  \mathcal{L}_p(\mathbb{P}_\text{nl} | \{z_l\}) 
		\nonumber\\
		&& 
		\qquad\quad
		= \min_{\mathbb{P}_\text{lin}} \left[ -2 \log  \mathcal{L}_c(\mathbb{P}_\text{lin} | \{z_l\}, \mathbb{P}_\text{nl})
		\right] .
		\label{eq:logLp}
	\end{eqnarray}
	In this manner, non-linear optimization is required to be carried out on the smaller subset of parameters $\mathbb{P}_\text{nl}$, while $\mathbb{P}_\text{lin}$ are straightforwardly solved by means of linear algebra.
	
	\subsubsection{Statistical weights and model fitting}
	
	The last statement holds if $\sigma_z$ and $\sigma_\lambda$ are known. 
	While comparing different realizations of the same history suggests the typical dispersion $\sigma_z \simeq 2-3 \times 10^{-3}$ (about twice the measurement noise), 
	we could not get a reliable estimate of the statistical fluctuations in $\lambda$ from data alone. 
	
	Under the assumption of normality on the error terms, each of the sums in (\ref{eq:logLc}) is $\chi^2$-distributed, with expectation values equal to the corresponding degrees of freedom, $\kappa_z = {\#\{z_l\}}-{\#\{A_j, \tilde{z}_{j,k}\}} $ and $\kappa_\lambda = \#\{\lambda_j\} - \#\{\beta, \tilde{\lambda_j}\}$, where $\#$ denotes the cardinality. 
	Then, 
	\begin{equation}
		\begin{aligned}
			\hat{\sigma}_z^2 = & {\frac{1}{\kappa_z} \sum_{j,k,l}\epsilon_{j,k,l}^2 }
			\; , \\
			\hat{\sigma}_\lambda^2 = & { \frac{1}{\kappa_\lambda} \sum_{j,k}\frac{\xi_{j,k}^2}{1 + (\sigma_\lambda e^{\lambda_j})^{-2}} }  
			\; , 
		\end{aligned}
		\label{eq:sigmaest}
	\end{equation}
	are unbiased estimators of the corresponding variances, conditional on the set  $\mathbb{P}_\text{nl}$ and the particular choice of $\sigma_\lambda$ used for fitting $\mathbb{P}_\text{lin}$.
	Thus, they are not to be taken as \emph{true} variance estimates, but rather as a means for assessing the relative goodness of fit of each component.
	At the same time, the \emph{optimal} $\mathbb{P}_\text{nl}$ and $\mathbb{P}_\text{lin}$ depend on $\sigma_\lambda$ and $\sigma_z$.
	Therefore, an iterative method was devised to estimate both  $\sigma_z$ and $\sigma_\lambda$, along with $\mathbb{P}_\text{nl}$ and $\mathbb{P}_\text{lin}$, in a self-consistent way:
	
	\begin{enumerate}
		\item Assign the initial values $\sigma_z = 2.5 \times 10^{-3}$ (based on raw data) and $\sigma_\lambda = \sigma_z$. We state these values for reference, but the algorithm converged to the same results regardless of the initial $\sigma_\lambda$.
		\item Find the maximum-likelihood estimates
		\begin{equation}
			\begin{aligned}
				\hat{\mathbb{P}}_\text{nl} =& \argmin_{\mathbb{P}_\text{nl}} \left[  -2 \log  \mathcal{L}_p(\mathbb{P}_\text{nl} | \{z_l\}) \right], \\
				\hat{\mathbb{P}}_\text{lin} =& \argmin_{\mathbb{P}_\text{lin}} \left[ -2 \log  \mathcal{L}_c(\mathbb{P}_\text{lin} | \{z_l\},\hat{\mathbb{P}}_\text{nl}) \right]
			\end{aligned}
		\end{equation}
		using weighted ordinary least squares (for ${\mathbb{P}}_\text{lin}$) nested within a general purpose optimization scheme (for ${\mathbb{P}}_\text{nl}$, see details below).
		\item Recalculate $\sigma_\lambda$ and $\sigma_z$ using equations (\ref{eq:sigmaest}) for the new parameters $\hat{\mathbb{P}}_\text{nl}$ and $\hat{\mathbb{P}}_\text{lin}$.
		\item If the last relative change in the ratio $\sigma_z^2 / \sigma_\lambda^2 > 10^{-5}$, repeat from step 2; else, convergence is assumed.
	\end{enumerate}
	
	All optimization was carried out on R, a free software environment for statistical computing~\cite{Rsoftware}.
	Linear model fitting was performed using the standard \texttt{lm} function.
	For non-linear optimization, a constrained quasi-Newton method was applied, as implemented under the name \texttt{L-BFGS-B} in the \texttt{optim} function.
	Bounds on $H_j^c$ were set to avoid problems associated with singularities ($0 \leq H_j^c \leq \min H_{j,k} - 10^{-3}~\text{Oe})$, while trivial bounds were chosen for other parameters ($0 \leq \alpha_j \leq 2$, $-10 \leq \lambda_{j,k} \leq 20$).
	
	\subsection{Parameter error estimation}
	
	While the concept of \emph{degrees of freedom} is unambiguous in the context of linear regressions, it becomes fuzzy in non-linear model fitting.
	This, in turn, makes error estimation a complicated task.
	We must note, however, that the final estimate $\hat{\sigma}_z = 2.9\times 10^{-3}$ is within our initial expectation, based on the typical dispersion of $\chi'(N)$ between independent realizations of the same history.
	Hence, we can arguably assume that the total \emph{effective} degrees of freedom lie somewhere between $\kappa_z + \kappa_\lambda$ and $\kappa_z + \kappa_\lambda - \#{\mathbb{P}}_\text{nl}$, which are still large enough  (i.e., between 299 and 335) to take advantage of the asymptotic properties of the log-likelihood function.
	
	The variance-covariance matrix of the parameters in $\mathbb{P}_\text{nl}$, $\Sigma_\text{nl}$, can now be estimated from the Hessian of the log-likelihood function,
	\begin{equation}
		\mathcal{H}_{p,p'} = \left. \frac{\partial^2 \log \mathcal{L}_p}{\partial p ~ \partial p'} \right|_{\hat{\mathbb{P}}_\text{nl}},
	\end{equation}
	as $\Sigma_\text{nl}(p,p') = -(\mathcal{H}^{-1})_{p,p'}$.
	Then, the corresponding estimated errors are $\sigma_p = \sqrt{\Sigma_\text{nl}(p,p)}$.
	
	In order to estimate errors for the parameters in $\mathbb{P}_\text{lin}$, it is necessary to consider both the errors in the linear fits as well as the errors inherited from $\mathbb{P}_\text{nl}$, on which $\mathbb{\hat P}_\text{lin}$ depends.
	The former can be straightforwardly evaluated using the usual expression for the variance-covariance matrix of linear model coefficients,
	\begin{equation}
		\Sigma_\text{lin,1}(p,p') = (M^{T} ~ \Sigma_\text{obs} ~ M)^{-1},
	\end{equation}
	where $M$ is the regressor (or design) matrix and  $\Sigma_\text{obs}$ is the variance-covariance matrix of the observations (which include both $\{z_l\}$ and $\{\lambda_{j,k}\}$).
	The second contribution can be calculated using the error propagation formula,
	\begin{equation}
		\Sigma_\text{lin,2}(p,p') = \sum_{q,q' \in \mathbb{P}_\text{nl} } \frac{\partial p}{\partial q} \frac{\partial p'}{\partial q'} \Sigma_\text{nl}(q,q').
	\end{equation}
	Finally, $\sigma_p = [\Sigma_\text{lin,1}(p,p) + \Sigma_\text{lin,2}(p,p)]^{1/2}$ yields an error estimate for these parameters.
	
	\section{Fitted models and scaling}
	
\begin{figure*}[h!]
		\begin{minipage}[t]{85mm}
			\centering
			\includegraphics[width=8.5cm]{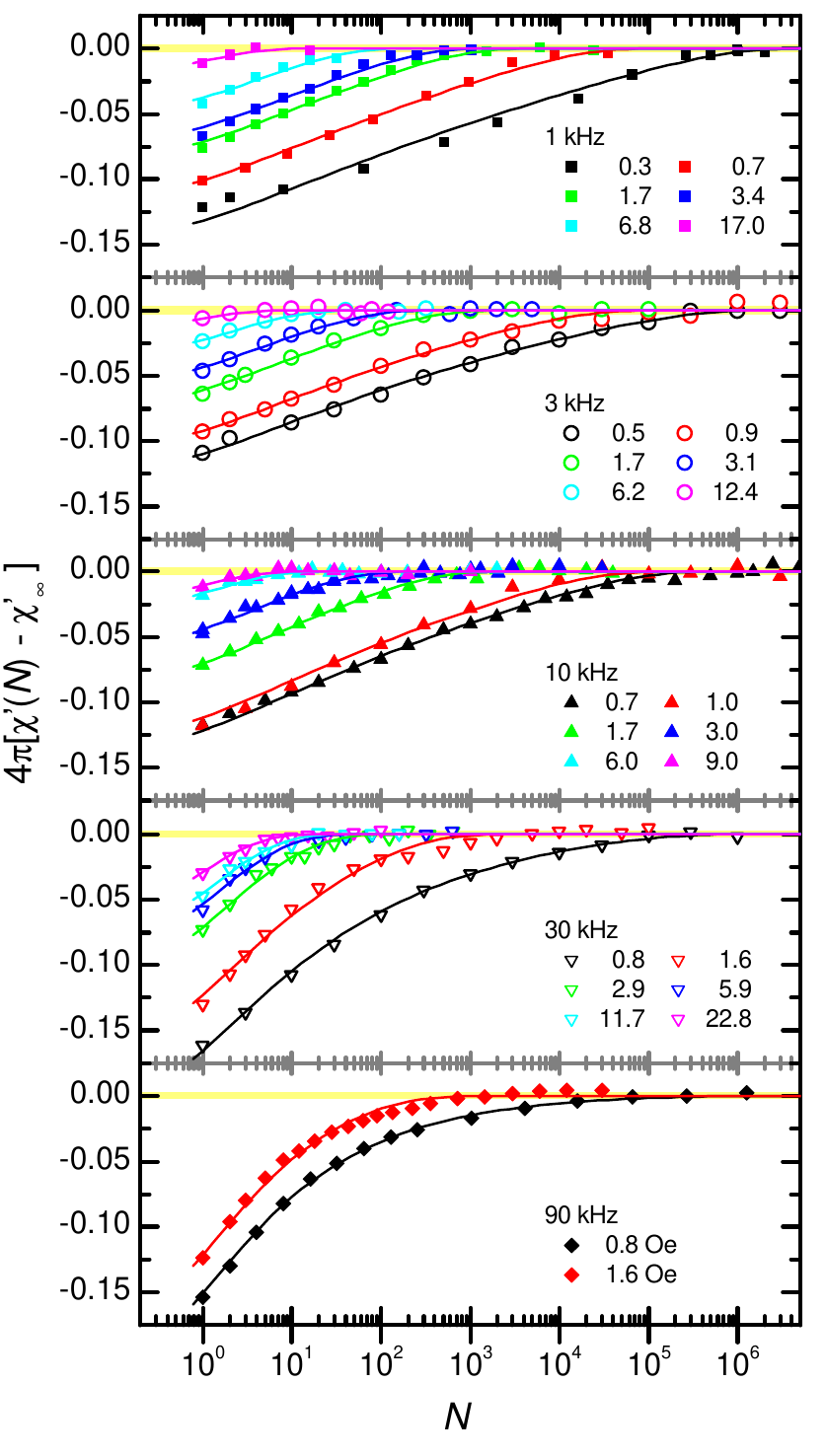}
			\caption{Measured $\chi'(N)$ (symbols) and fitted model curves (lines) for each series. Error bars are not displayed, as they would be about the size of the symbol. Instead, a yellow box of width equal to 2$\sigma$ centered about $0$ is shown.}
			\label{fig:data}
		\end{minipage}\hfill\begin{minipage}[t]{85mm}
		\centering
		\includegraphics[width=8.5cm]{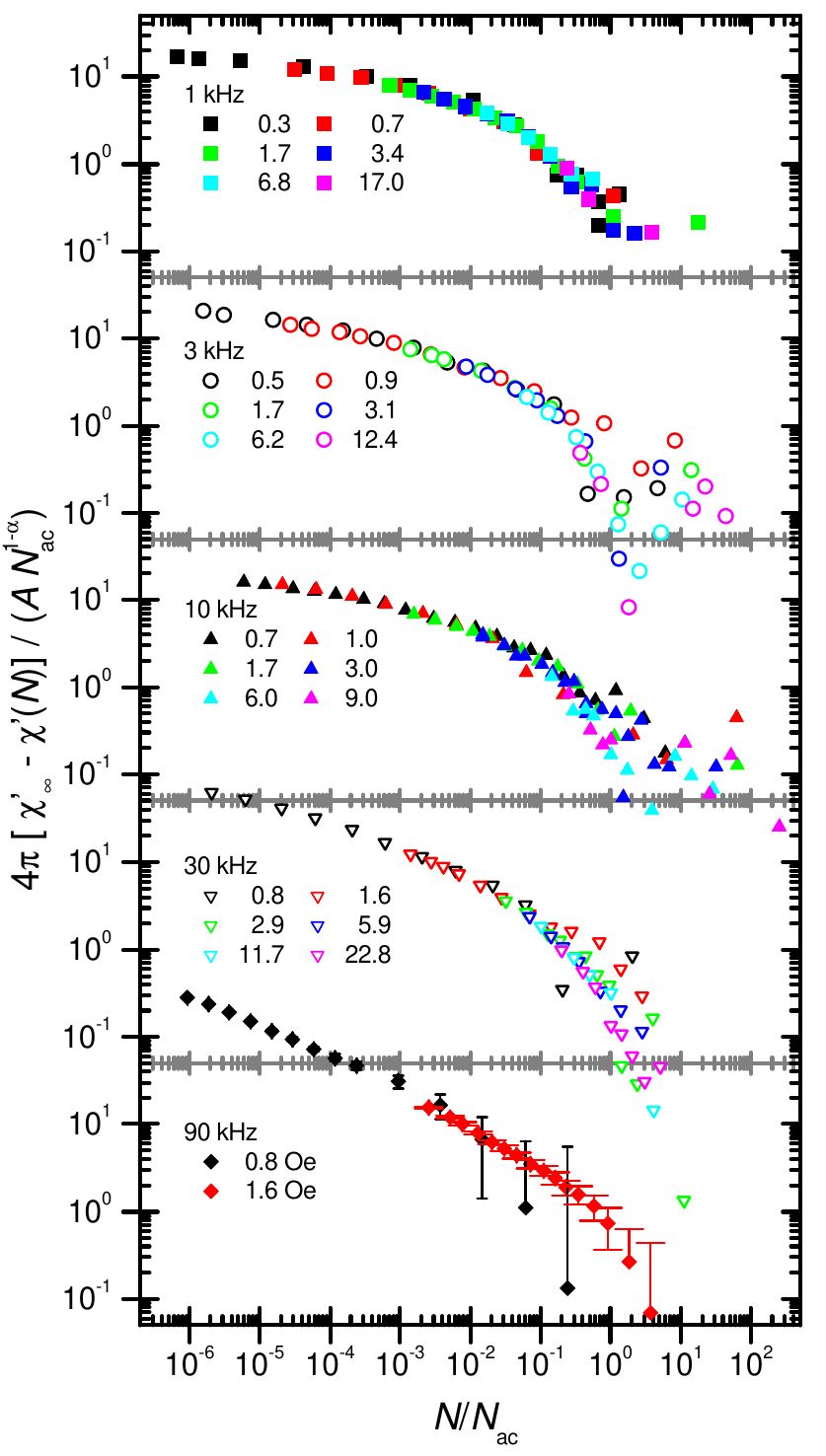}
		\caption{
			Scaled $\chi'(N/N_{\text{ac}})$ collapsing to a universal function. 
			As an example, scaled error bars are shown for $f_{\text{sh}} = 90~\text{kHz}$ only. }
		\label{fig:scaling}
	\end{minipage}
\end{figure*}

	Figure \ref{fig:data} shows the evolution of the linear response with the number of applied pulses $N$ tending to the stationary response $\chi'_{\infty } (H_\text{sh}, f_\text{sh})$. Each panel displays the evolution measured at each shaking frequency at different shaking amplitudes. Continuous lines 
	are fits of the data with the function (\ref{eq:chiSumN}) using the  exponents $\alpha (f_{sh}) $ and prefactors $A (f_\text{sh})$ obtained with the above fitting procedure.
	
	As mentioned in the main text, Eq.~(\ref{eq:chiGamma}) implies a scaling of the experimental data plotted as a function of the variable $N/N_\text{ac}$, in the form
	\begin{equation}\label{eq:scaling}
		\frac{4\pi }{A N_\text{ac}^{1-\alpha } }(\chi'_{\infty }-\chi'(N)) \simeq\Gamma _{1-\alpha }\left( \frac{N}{N_\text{ac}}\right) 
		\; , 
	\end{equation}
	for $N\gg 1$. The scaling obtained for the data set corresponding to all the measured shaking frequencies is shown in Fig.~\ref{fig:scaling}. 
	For each $f_\text{sh}$ we are able to collapse all the data in a single curve, corresponding to the function $\Gamma_{1-\alpha} (N/N_\text{ac})$ 
	where the characteristic number of cycles $N_\text{ac} (H_\text{sh},f_\text{sh})$ and the exponent $\alpha (f_\text{sh})$ are those obtained with the procedure above described.

\bibliography{smref}